\documentstyle[11pt]{article}

\begin{document}

\begin{center}
{\Large \bf Action Principle for the Classical Dual
Electrodynamics\footnote{Work partially supported by CNPq and FAPESP}}
\end{center}

\begin{center}
{\large \bf PCR Cardoso de Mello$^{(1)}$, S Carneiro$^{(2)}$ and 
MC Nemes$^{(1)}$}
\end{center}

\begin{center}
$^{(1)}$Departamento de F\'{\i}sica, Universidade Federal de Minas
Gerais\\
CP702, 30000,
Belo Horizonte, MG, Brasil.
\end{center}

\begin{center}
$^{(2)}$Instituto de F\'{\i}sica, Universidade de S\~ao Paulo\\
CP66318, 05389-970, S\~ao Paulo, SP, Brasil.
\end{center}

\begin{abstract}
The purpose of this paper is to formulate an action principle which
allows for the construction of a classical lagrangean including both
electric and magnetic currents. The lagrangean is non-local and shown
to yield all the expected (local) equations for dual electrodynamics.
\vspace{0.5cm}
\end{abstract}

One of the oldest open problems in the theory of elementary particles
is that of the quantization of the electric charge. Although
apparently very simple, this experimental result has not yet found
theoretical explanation in the context of the standart model of
fundamental interactions.

In 1931, PAM Dirac$^{[1]}$ found an explanation for such quantization
based on the lack of symmetry of Maxwell's equations in what concerns
their source terms. The presence of magnetic currents in these
equations leads, at the quantal level, to the quantization of the
electric and magnetic charges.

Since the pioneer work of Dirac, other solutions to the problem have
been proposed in the context of unified theories, as GUT's$^{[2,3]}$
and Kaluza-Klein theories$^{[4,5]}$. However, all these proposals are
shown to be connected to the existence of solitonic magnetic
monopoles$^{[6-9]}$.

A great obstacle to the development of an electrodynamics with charges
and poles is the absence of an adequate lagrangean formulation. This
is intimately connected to the difficulty of constructing a regular
$4$-potential in all space-time. There have been several proposals to
circumvent this problem: the introduction of Dirac's string$^{[10]}$,
of the double-valued Wu-Yang potential$^{[11]}$, of the singular
Bollini-Giambiagi potential$^{[12]}$ and finally the use of non-local
wave functions, proposed by Cabbibo and Ferrari$^{[13]}$. However, a
lagrangean formulation which gives rise to the complete set of
electromagnetic equations, without necessity of any subsidiary
condition, is still lacking.

The main purpose of the present work is to show that a non-local
lagrangean can be constructed which gives a correct description of the
classical dual electrodynamics provided we postulate the following
variational principle: {\it the dynamics of the system
charge-field-monopole is such that the action presents a saddle point
which is a minimum with respect to variation of the usual degrees of
freedom and a maximum with respect to variation of the dual degrees of
freedom}.

Following such prescription we construct the lagrangean density

\begin{equation}
\label{1}
{\cal L} = {\cal L}_o^e + {\cal L}_o^g - \frac{1}{4} F_{\mu \nu}
F^{\mu \nu} - j_{\mu}{\cal A}^{\mu} + g_{\mu}\tilde{{\cal A}}^{\mu} 
\end{equation}

\noindent where $j_{\mu}$ e $g_{\mu}$ are the electric and magnetic
$4$-currents, respectively. Here we have introduced
the Cabibbo-Ferrari generalized field tensor$^{[13]}$

\begin{equation}
\label{2}
F^{\mu \nu} \equiv \partial^{\mu} A^{\nu} - \partial^{\nu} A^{\mu} -
\epsilon^{\mu \nu \alpha \beta} \; \partial_{\alpha} \tilde{A}_{\beta} 
\end{equation}

\noindent and the non-local potentials$^{[14]}$

\begin{equation}
\label{3}
{\cal A}^{\mu} = A^{\mu} + \frac{1}{2}
\epsilon^{\mu \gamma \alpha \beta} \int_P^x \partial_{\alpha}
\tilde{A}_{\beta} \; d\xi_{\gamma}
\end{equation}

\begin{equation}
\label{4}
\tilde{{\cal A}}^{\mu} =
\tilde{A}^{\mu} - \frac{1}{2} 
\epsilon^{\mu \gamma \alpha \beta} \int_{\tilde{P}}^x \partial_{\alpha}
A_{\beta} \; d\xi_{\gamma}
\end{equation}

\noindent with $P$ and $\tilde{P}$ defined, respectively, by the world
lines of the charge and pole.

The first two terms in (\ref{1}) correspond, respectively, to the free
lagrangeans of the electric and magnetic charges, so that the
Lagrange's function cor\-responding to (\ref{1}) is given by

\begin{equation}
\label{5}
L = L_e + L_g + L_{Maxwell}
\end{equation}

\noindent where

\begin{equation}
\label{6}
L_e = - m \;(1 - u^2)^{\frac{1}{2}} + e \;\vec{u} \cdot \vec{{\cal A}}
- e \; {\cal A}_0
\end{equation}

\begin{equation}
\label{7}
L_g = M \;(1 - v^2)^{\frac{1}{2}} - g \;\vec{v} \cdot
\vec{\tilde{{\cal A}}} + g \;\tilde{{\cal A}}_0
\end{equation}

\begin{equation}
\label{8}
L_{Maxwell} = - \frac{1}{4} \int d^3x \; F_{\mu \nu}F^{\mu \nu}
\end{equation}

\noindent Here $\vec{u}$ and $\vec{v}$ stand for the charge and pole
velocities, $m$ and $M$ standing for their masses. $e$ and $g$ are
their charge strenghts.

In the absence of monopoles we can, using the gauge freedom$^{[13]}$,
set $\tilde{A}^{\mu} = 0$ and our lagrangean reduces to the usual
lagrangean of electromagnetism. Also, in the absence of electric
charges, by setting $A^{\mu} = 0$ we get, apart from an overall sign,
the dual lagrangean

\begin{equation}
\label{9}
{\cal L} = {\cal L}_o - \frac{1}{4} \tilde{F}_{\mu \nu} \tilde{F}^{\mu
\nu} - g_{\mu} \tilde{A}^{\mu} 
\end{equation}

\noindent where 

\begin{equation}
\label{10}
\tilde{F}^{\mu \nu} = \frac{1}{2} \epsilon^{\mu \nu \alpha \beta}
F_{\alpha \beta}
\end{equation}

\noindent stands for the dual of the field tensor.

From (\ref{2})--(\ref{4}) and (\ref{10}) we can show the validity of
the relations$^{[17]}$

\begin{equation}
\label{11}
F^{\mu \nu} = \partial^{\mu} {\cal A}^{\nu} - \partial^{\nu}
{\cal A}^{\mu}
\end{equation}

\begin{equation}
\label{12}
\tilde{F}^{\mu \nu} = \partial^{\mu} \tilde{{\cal A}}^{\nu} - \partial^{\nu}
\tilde{{\cal A}}^{\mu}
\end{equation}

Variations of the local potentials -- for given fixed particles's
world lines -- lead to variations of the non-local ones. Using
(\ref{11}), (\ref{12}) and the identity

\begin{equation}
\label{13}
F_{\mu \nu}F^{\mu \nu} = - \tilde{F}_{\mu \nu}\tilde{F}^{\mu \nu}
\end{equation}

\noindent it is a simple matter to check that the extremum condition
for the action under such variations leads to Euler-Lagrange equations
which correspond to the expected generalized Maxwell's equations

\begin{equation}
\label{14}
\partial_{\beta} F^{\alpha \beta} = - j^{\alpha}
\end{equation}

\begin{equation}
\label{15}
\partial_{\beta} \tilde{F}^{\alpha \beta} = - g^{\alpha}
\end{equation}

Variations with respect to the coordinates of the charge and pole give

\begin{equation}
\label{17}
m \frac{dU^{\alpha}}{d\tau} = e \; (\partial^{\alpha}
{\cal A}^{\beta} - \partial^{\beta} {\cal A}^{\alpha}) \; U_{\beta}
\end{equation}

\begin{equation}
\label{18}
M \frac{dV^{\alpha}}{d\tau} = g \; (\partial^{\alpha}
\tilde{{\cal A}}^{\beta} - \partial^{\beta} \tilde{{\cal A}}^{\alpha})
\; V_{\beta}
\end{equation}
 
\noindent where $U^{\mu}$ and $V^{\mu}$ stand for the $4$-velocities
of the charge and pole and $\tau$ is their proper time. Here, the
derivatives of the potentials are taken along the world lines of the 
particles. Thus, using again (\ref{11}) and (\ref{12}), we obtain the
correct Lorentz's equations

\begin{equation}
\label{19}
m \frac{dU^{\alpha}}{d\tau} = e F^{\alpha \beta} \; U_{\beta}
\end{equation}

\begin{equation}
\label{20}
M \frac{dV^{\alpha}}{d\tau} = g
\tilde{F}^{\alpha \beta} \; V_{\beta}
\end{equation}

We can see that the proposed lagrangean, although non-local, leads to
all desired local equations of motion. Using the field equations it is
possible to show that a change of the paths of integration in
(\ref{3}) and (\ref{4}) corresponds to a gauge transformation of the 
non-local potentials. Because this result, it is not necessary to
consider variations of these paths to obtain the particles's equations.

It is important to note that (\ref{11}) and (\ref{12}) do not imply
into the homogeneity of (\ref{14}) and (\ref{15}). It is due to the
fact that the non-local potentials are not regular, do not obeing the
Euler condition. In other words,

\begin{equation}
\label{16}
(\partial^{\mu} \partial^{\nu} - \partial^{\nu}
\partial^{\mu}) \; {\cal A}^{\alpha} \neq 0
\end{equation}

\noindent and the same for $\tilde{{\cal A}}^{\alpha}$.

The irregular character of ${\cal A}^{\mu}$, as a function of $x$ and
$P$, is evident once one examines expression (\ref{3}) in the case of
a magnetic monopole at rest in the origin. Any path which goes through
the origin turns the integral into a divergent one over a
semi-infinite line. Such singularity are essential, since they come
from the intersection between the charge's world line and the
monopole's one, and are already contained in the equations of motion
derived from the lagrangean. In fact, Lorentz's equation (\ref{19})
allows the charge to come indefinitely close to the monopole, over the
line connecting them. But when the superposition occurs, the second
member of this equation becomes singular, unless the relative velocity
between charge and pole goes to zero. We note however that, whatever
the charge's trajectory might be, the singularity of ${\cal A}^{\mu}$
will always lie in a $4$-hemisphere opposed to that of charge's
motion. This discussion is valid also for the dual non-local potential.

Let us consider the dual transformation

\begin{equation}
\label{21}
A^{\mu} \rightarrow - \tilde{A}^{\mu}
\end{equation}

\begin{equation}
\label{22}
\tilde{A}^{\mu} \rightarrow A^{\mu}
\end{equation}

\begin{equation}
\label{23}
j^{\mu} \rightarrow - g^{\mu}
\end{equation}

\begin{equation}
\label{24}
g^{\mu} \rightarrow j^{\mu}
\end{equation}

\noindent together with $m \leftrightarrow M$. The lagrangean and
action will change sign. As the equations of motion do not depend on
the overall sign of these quantities, we can say that the theory
remains invariant under such dual transformation.

The sign difference between the free lagrangeans of the electric and
magnetic charges (cf. (\ref{6}) and (\ref{7})) may give the impression
that the monopole would appear as a particle with negative
energy$^{[15]}$. This would be of course unacceptable at the classical
level. It is possible to see that this is not the case by calculating
the total conserved energy of the system charge-field- monopole, using
the equations of motion obtained from the lagrangean.

This result apparently contradicts the hamiltonian formulation of the
theory. However, the dual simetry and the very form of the lagrangean,
obeing a saddle principle, lead us to a hamiltonian formulation which
is internally consistent with the theory. In fact, the dual
transformation (\ref{21})--(\ref{24}), under which $L_e \rightarrow
-L_g$ and $S_e \rightarrow -S_g$, transforms the momentum and
hamiltonian of the charge

\begin{equation}
\label{25}
\vec{p}_e \equiv \frac{\partial S_e}{\partial \vec{r}} =
\frac{\partial L_e}{\partial \vec{u}} = \frac{m \vec{u}}{(1 -
u^2)^{\frac{1}{2}}} + e \vec{{\cal A}} 
\end{equation}

\begin{equation}
\label{26}
{\cal H}_e \equiv - \frac{\partial S_e}{\partial t} = \frac{\partial
L_e}{\partial \vec{u}} \cdot \vec{u} - L_e = [m^2 + (\vec{p}_e - e
\vec{{\cal A}})^2]^{\frac{1}{2}} + e  {\cal A}_0 
\end{equation}

\noindent into the momentum and hamiltonian of the pole

\begin{equation}
\label{27}
\vec{p}_g = - \frac{\partial S_g}{\partial \vec{r}} = - \frac{\partial
L_g}{\partial \vec{v}} = \frac{M \vec{v}}{(1 - v^2)^{\frac{1}{2}}} + g
\vec{ \tilde{{\cal A}}} 
\end{equation}

\begin{equation}
\label{28}
{\cal H}_g = \frac{\partial S_g}{\partial t} = - \left( \frac{\partial
L_g}{\partial \vec{v}} \cdot \vec{v} - L_g \right) = [M^2 + (\vec{p}_g
- g  \vec{\tilde{{\cal A}}})^2]^{\frac{1}{2}} + g  \tilde{{\cal A}}_0 
\end{equation}

The above expressions respect the canonical form of Hamilton's
equations, since the latter remain invariant under a simultaneous
change of sign of $\vec{p}$ and ${\cal H}$. It is also simple to show
that (\ref{28}) corresponds to the correct time evolution generator
for the monopole.

We should like to remark that our formulation does not induce any 
mo\-dification for the particles's equations of motion in the
gravitational field. 

Usually the action for a mass $m$ particle subject to this field is given by

\begin{equation}
\label{29}
S = - m \int\; ds
\end{equation}

\noindent with

\begin{equation}
\label{30}
ds^2 = g_{\mu \nu} \; dx^{\mu} \; dx^{\nu}
\end{equation}

In the case of a magnetic charge, however, one should consider the
action in the form

\begin{equation}
\label{31}
S = M \int\; ds
\end{equation}

\noindent so that it presents, contrary to (\ref{29}), a maximum and
so that the Lagrange's function reduces to the correct one
(cf. (\ref{7})) in the absence of the gravitational field.

Since the equations of motion are given by $\delta S = 0$ the sign
difference between (\ref{29}) and (\ref{31}) will not matter. This is
in complete agreement with the Equivalence Principle. Besides, the
positive definite character of the energy of the monopole guarantees
that it will play the same role as any other particle in what concerns
the generation of gravitational field.

In conclusion, we have proposed an action principle which allows for
the construction of a non-local classical lagrangean which yields all
the equations of electromagnetism with charges and monopoles, without
having to resort to additional restrictions or constraints on the
dynamics of the particles. 

The quantization of the theory remains a challenging open problem. The
same can be said of its non-abelian extension. In the same way that
magnetic monopoles can be obtained as solitons of non-abelian
theories, we can think that electric charges would be given as
topological solutions of dual theories to the first$^{[19-21]}$. This
possibility suggests a unified description of electric and magnetic
charges as configurations of bosonic scalar and vector fields. The
difficulty lies, however, in the lack of a self-dual lagrangean which
contains at the same time the bosonic fields and their respective
duals, like in (\ref{1}). The introduction of non-local potentials
(non-abelian) may be a way to the construction of a saddle point
lagrangean formulation.


\begin{thebibliography}{99}
\bibitem{14} Around three decades ago, Rohrlich$^{[15]}$ introduced
similar non-local potentials, making use of a parametrization
suggested by DeWitt$^{[16]}$ for trecking non-local wave functions
like the one used by Cabibbo and Ferrari$^{[13]}$. In his work,
however, he arrives at different results and conclusions.
\bibitem{17} The covariant derivative of the line integral of a tensor
$\Lambda^{\mu...\nu}$ with \-res\-pect to the observation point
$x^{\mu}$, extremum of the path of integration $P$, along this path,
is defined as$^{[18]}$

\begin{equation}       
\label{32}
\partial^{\mu} \int_P^x \Lambda^{\alpha...\nu}(\xi) \; d\xi_{\nu}
\equiv
\lim_{\begin{array}{l}
\scriptsize
dx_{\mu} \rightarrow 0
\end{array}} \;
\frac{\left[\int_{P'}\Lambda^{\alpha...\nu}(\xi)\;d\xi_{\nu} -
\int_P\Lambda^{\alpha...\nu}(\xi)\;d\xi_{\nu}\right]}{dx_{\mu}}
\nonumber
\end{equation}

\noindent where $P'$ is obtained from $P$ by adding to this one an
extension $dx^{\mu}$ in the $x^{\mu}$ direction. Thus if the
derivation is to be performed in a direction which should be
orthogonal to that of the integration, the diference in brackets
vanishes and so does the derivative. On the other hand, if we derive
in the same direction of the integration, we obtain, according to the
Fundamental Theorem of Calculus, $\Lambda^{\alpha...\mu}(x)$. Thus we have

\begin{equation}
\label{33}
\partial^{\mu} \int_P^x \Lambda^{\alpha...\nu}(\xi) \; d\xi_{\nu} =
\Lambda^{\alpha...\nu}(x) \; \delta^{\mu}_{\nu} =
\Lambda^{\alpha...\mu}(x) \nonumber
\end{equation}
\end{thebibliography}
\end{document}